\documentclass[aps,prd,reprint,showkeys,noshowpacs,longbibliography,floatfix,amsmath,amssymb,nofootinbib,noshowpacs,superscriptaddress,notitlepage,preprintnumbers]{revtex4-2}

\usepackage{xcolor}
\usepackage{graphicx}% Include figure files
\usepackage{dcolumn}% Align table columns on decimal point
\usepackage{bm}% bold math
\usepackage{hyperref}% add hypertext capabilities
\usepackage[compat=1.0.0]{tikz-feynman}
\begin{document}

\preprint{ADP-26-16/T1313, DESY-26-133}

\title{Bjorken sum rule down to the elastic region from the lattice}

\author{J.~A.~Crawford}
\affiliation{
CSSM, Department of Physics, Adelaide University, Adelaide SA, 5005, Australia}
\affiliation{
Department of Physics, Bergische Universit\"at Wuppertal, 42119 Wuppertal, Germany}
\author{A.~Hannaford-Gunn}
\affiliation{
CSSM, Department of Physics, Adelaide University, Adelaide SA, 5005, Australia}
\author{R.~Horsley}
\affiliation{School of Physics and Astronomy, University of Edinburgh, Edinburgh, EH9 3FD, UK}
\author{P.~E.~L~Rakow}
\affiliation{Theoretical Physics Division, Department of Mathematical Sciences, University of Liverpool, Liverpool L69 3BX, UK}
\author{T.~G.~Schar}
\affiliation{
CSSM, Department of Physics, Adelaide University, Adelaide SA, 5005, Australia}
\author{G.~Schierholz}
\affiliation{Deutsches Elektronen-Synchrotron DESY, Notkestr. 85, 22607 Hamburg, Germany}
\author{H.~St\"uben}
\affiliation{Universit\"at Hamburg, Regionales Rechenzentrum, 20146 Hamburg, Germany}
\author{R.~D.~Young}
\affiliation{
CSSM, Department of Physics, Adelaide University, Adelaide SA, 5005, Australia}
\author{J.~M.~Zanotti}
\affiliation{
CSSM, Department of Physics, Adelaide University, Adelaide SA, 5005, Australia}

\collaboration{QCDSF Collaboration}

\date{\today}% It is always \today, today,

\begin{abstract}
We present a lattice determination of the Bjorken sum rule via a Feynman-Hellmann calculation of the polarised forward Compton amplitude. We evaluate the lowest moment of the nucleon's polarised structure function, $g_1$, which we calculate over the range $0.5\lesssim Q^2 \lesssim 8$ GeV$^2$, covering both non-perturbative and perturbative regions. We compare our calculated moments with the Bjorken sum rule, and discuss implications for the calculation of higher-twist effects, as well as evaluations of $\alpha_s(Q^2)$ from hadronic quantities to complement phenomenological and other lattice methods.
\end{abstract}

\maketitle

%\tableofcontents

\section{Introduction}
\label{sec: Intro}
The polarised structure function $g_1$, accessible via polarised deep-inelastic scattering experiments, is central to our understanding of the spin structure of the nucleon. At the partonic level, the lowest moment of $g_1$ is related to the fraction of the total hadron spin carried by the partons,
\begin{equation}
    \int_0^1dx\,g_1(x,Q^2) \xrightarrow{Q^2\to\infty} \frac{1}{2}\sum_qe_q^2[\Delta q(Q^2) + \Delta \overline{q}(Q^2)],
\end{equation}
where $e_q$ is the electric charge and $\Delta q$ $(\Delta \overline{q})$ is the contribution to the spin from the (anti-)parton of flavour $q$. Exploiting iso-spin symmetry, the Bjorken sum rule \cite{Bjorken1970} is defined from the difference between the lowest moment of the proton and neutron $g_1$ structure functions,
\begin{equation}
    \label{eq: Bjorken sum rule}
    \Gamma_1^{p-n}(Q^2) = \int_0^1dx\, [g^p_1(x,Q^2)-g_1^n(x,Q^2)],
\end{equation}
which in the quark-parton model is equal to $g_A^{u-d}/6$, with $g_A^{u-d}$ being the isovector axial charge. As the Bjorken sum rule is a flavour non-singlet quantity, radiative corrections can be calculated to a high degree of precision in perturbative QCD, therefore offering a robust avenue for precision studies of QCD dynamics. The Bjorken sum rule has been experimentally validated, most precisely by E155 at $Q^2=5$ GeV$^2$ \cite{Jlab2000}. The $Q^2$ dependence of the sum rule has been mapped out at Jefferson Lab \cite{Deur:2004ti,Jlab2008, Jlab2014, CLAS2017}, and shows agreement with theoretical predictions down to $1$ GeV$^2$. In the very low $Q^2$ range, recent experiments \cite{CLAS2018, CLAS:2021apd, CLAS2024} have mapped the $Q^2$ dependence of the Bjorken sum rule to as low as $0.01$ GeV$^2$. In this $Q^2$ range, the $Q^2$ dependence of the integral in Equation \eqref{eq: Bjorken sum rule} is instead governed by the Gerasimov-Drell-Hearn (GDH) sum rule \cite{Gerasimov:1965et, Hosoda:1966rpz, Drell1966}. The transition between these two regimes is still not well understood.
\newline

Lattice QCD offers a first-principles approach to the calculation of matrix elements of local operators. Calculations of the isovector axial charge have reached the sub-percent level \cite{Chang2018, RQCD2019, Bali2023, QCDSFUKQCDCSSM2023, Jang2023, Djukanovic2024, Alexandrou2024}, which is required for the Bjorken sum rule at the leading-twist level. Higher-twist contributions are suppressed by inverse powers of $Q^2$ and therefore become relevant at low $Q^2$. However, direct lattice calculations of higher-twist matrix elements are challenging. On the lattice, operators can mix with others transforming in the same irreducible representation of the hypercubic group \cite{Kronfeld1985,Gockeler1996,Gockeler2001}, including lower-dimensional operators that induce power-divergent mixing. Twist-three matrix elements remain tractable in suitably chosen channels \cite{Gockeler2000,Gockeler2005,Burger2021,Crawford2024,Gao2026}; however, corresponding calculations of twist-four matrix elements and beyond have not been pursued, as the operator-mixing structure becomes substantially more complicated and is not fully known. These sub-leading twist contributions offer unique insight into hadron structure, going beyond the parton model by probing quark-gluon correlations \cite{Shuryak1981} and are essential for understanding the internal forces that bind quarks into hadrons. Alternative methods for calculating these higher-twist matrix elements are hence highly desirable. 
\newline 

Beyond hadron structure implications, the Bjorken sum rule offers an excellent avenue for a non-perturbative definition of the strong coupling $\alpha_s$ across all momentum scales. As the Bjorken sum rule is a non-singlet quantity, at leading order in QCD it has vanishing anomalous dimension. Additionally, the non-singlet Wilson coefficient is known to next-to-next-to-next leading order (N$^3$LO) \cite{Larin1991}. These two facts make the Bjorken sum rule, as well as the Gross-Llewellyn Smith sum rule \cite{Gross1969}, ideal candidates for determining $\alpha_s$ \cite{Kim1998, Deur2005}, and for a non-perturbative definition of $\alpha_s$ via an effective charge approach analogous to QED (see Refs. \cite{Deur2016, Deur2023} for reviews). First principles calculations of $\alpha_s$ via these sum rules are highly desirable as they illuminate the boundaries between perturbative and non-perturbative regimes, offer insights into confinement  and provide an alternative determination of $\alpha_s$ based on a hadronic observable.
\newline

In this work, we use the Feynman–Hellmann technique to determine the Bjorken sum rule across a range of hard scales, $Q^2$, and compare with predictions from the parton model and perturbative QCD. The technique has proven to be a valuable approach for evaluating the Compton amplitude in lattice QCD \cite{Chambers2017}, with applications to moments of unpolarised nucleon structure functions \cite{Can2020,QCDSF2022}, generalised parton distributions \cite{QCDSF2021,Hannaford-Gunn2024}, and the Gross–Llewellyn Smith sum rule \cite{Can2025}. Alongside direct four-point-function calculations of the Compton amplitude \cite{Feng:2020zdc,Ma:2023kfr}, it provides a complementary framework for evaluating moments of structure functions across a broad range of probing momenta, enabling studies of power corrections and access to higher-twist matrix elements without the complications of power-divergent operator mixing. Preliminary investigations of spin-dependent structure functions using this approach were reported in Ref.~\cite{Can2022}.
\newline

The rest of the paper is structured as follows: In Section \ref{sec: Compton Amplitude} we present the tensor decomposition of the Compton amplitude and relate the polarised Compton structure function to the nucleon structure function $g_1$. Our Feynman-Hellmann approach is explained in Section \ref{sec: FH Approach}, followed by details of our lattice calculation and analysis in Section \ref{sec: Calc details}. We present and discuss our results in Section \ref{sec: Results}, before summarising and offering some concluding remarks in Section \ref{sec: Summary}. 
\newline

\section{The Spin-Dependent Compton Amplitude} 
\label{sec: Compton Amplitude}
The forward Compton amplitude is the time-ordered product of two vector currents inserted between two nucleon states,
\begin{equation}
    \label{eq: Compton amplitude}
    [T_{\mu\nu}(p,q)]_{\sigma'\sigma}=i\int d^4z\,e^{iq\cdot z}\langle p,\sigma' |\mathcal{T}\{\mathcal{J}_\mu(z)\mathcal{J}_\nu(0) \} |p,\sigma \rangle,
\end{equation}
where $p$ $(q)$ is the nucleon (current) momentum, $\sigma',\sigma$ are the outgoing and incoming nucleon spins and the vector current is defined as
\begin{equation}
    \label{eq: vec current}
    \mathcal{J}_\mu(z) = \sum_f \mathcal{Q}_f \overline{\psi}_f(z)\gamma_\mu \psi_f(z),
\end{equation}
with $\mathcal{Q}_f$ being the electric charge of the quark flavour $f$. To access the structure functions, we consider the Lorentz decomposition of the Compton amplitude \cite{Ji:1993ey,Manohar1992} in Minkowski space,
\begin{widetext}
\begin{multline}
    \label{eq: Tensor decomposition}
    [T_{\mu\nu}(p,q)]_{\sigma'\sigma} = \bigg( -g_{\mu\nu} + \frac{q_\mu q_\nu}{q^2}\bigg)\mathcal{F}_1(\omega,Q^2) + \bigg(p_\mu - \frac{p\cdot q}{q^2}q_\mu \bigg)\bigg(p_\nu - \frac{p\cdot q}{q^2}q_\nu \bigg)\frac{\mathcal{F}_2(\omega,Q^2)}{p\cdot q} \\
    + \frac{i}{p\cdot q}\varepsilon_{\mu\nu\alpha\beta}q^\alpha s_{\sigma'\sigma}^\beta \mathcal{G}_1(\omega,Q^2) - \frac{i}{(p\cdot q)^2}\varepsilon_{\mu\nu\alpha\beta}q^\alpha(p\cdot q s^\beta_{\sigma'\sigma} - s_{\sigma'\sigma}\cdot q p^\beta)\mathcal{G}_2(\omega,Q^2)
\end{multline}
\end{widetext}
where $\sigma$ $(\sigma')$ is the spin index of the incoming (outgoing) nucleon, $\varepsilon_{0123}=+1$, $\omega = 2p\cdot q/Q^2$, $Q^2=-q^2$ and the nucleon structure functions are $\mathcal{F}_{1,2}$ and $\mathcal{G}_{1,2}$ for the unpolarised and polarised cases respectively (the normalisations have been chosen to match the scaling functions of the Particle Data Group \cite{PDG2024}). $s^\mu_{\sigma'\sigma}$ is the spin matrix of the nucleon axial current bilinear, defined as
\begin{equation}
    s^\mu_{\sigma'\sigma}=\frac{1}{2}\overline{u}(p,\sigma')\gamma^\mu\gamma^5 u(p,\sigma),
\end{equation}
and satisfies $s^2=-m^2$. In order to access the polarised structure functions, we must consider off-diagonal components of the Compton amplitude, $T_{\mu\nu}$ with $\mu\neq\nu$. From analytic properties of the Compton tensor, crossing-symmetry relations and the optical theorem, the connection between the Compton structure function and the physical structure function is given by a dispersion relation,
\begin{equation}
\label{eq: dispersion relation}
    \mathcal{G}_1(\omega,Q^2) = 4\omega \int_0^1dx\, \frac{g_1(x,Q^2)}{1-x^2\omega^2}.
\end{equation}
On the lattice, the Wick rotation restricts access to the unphysical region $|\omega|<1$ \cite{Can2020}. In this region, the denominator may be expanded in a geometric series, allowing the Compton structure function to be expressed in terms of Cornwall--Norton moments of $g_1$:
\begin{equation}
    \label{eq: moment expansion}
    \frac{\mathcal{G}_1(\omega,Q^2)}{\omega}=4\sum_{n=1}^\infty \omega^{2n-2} \Tilde{M}_{2n-1}^{(1)}(Q^2).
\end{equation}
The moments of $g_1$ are defined as
\begin{equation}
\label{eq: g1 moment expansion}
    \Tilde{M}_{2n-1}^{(1)}(Q^2)=\int_0^1dx\,x^{2n-2}g_1(x,Q^2),
\end{equation}
for $n=1,2,3,\ldots$. The $n=1$ moment gives direct access to the Bjorken sum rule
\begin{equation}
    \Gamma_1^{p-n}(Q^2) = \Tilde{M}^{(1)}_{1,p}(Q^2) - \Tilde{M}^{(1)}_{1,n}(Q^2).
\end{equation}

\section{Feynman-Hellmann Approach}
\label{sec: FH Approach}
The Feynman–Hellmann approach provides an alternative formulation of the Compton amplitude in terms of two-point correlation functions, avoiding the need for an explicit four-point-function calculation.
\newline

We proceed by introducing two perturbing fields to the bare QCD action,
\begin{multline}
\label{eq: action perturbation}
    S(\bm{\lambda})=S_{0}+\lambda_1\int d^4z\,(e^{iq\cdot z}+e^{-i q\cdot z})\mathcal{J}_\mu(z) \\
    -i\lambda_2 \int d^4y\,(e^{iq\cdot y} - e^{-iq\cdot y})\mathcal{J}_\nu(y),
\end{multline}
where $S_0$ is the unperturbed action and $\bm{\lambda}=(\lambda_1,\lambda_2)$ denotes the coupling strength of the external perturbing fields and the quark fields. The current $\mathcal{J}_\mu(z)=Z_V \overline{q}_f(z)\gamma_\mu q_f(z)$ is the renormalised vector currents coupling to the quark fields of flavour $f$ along the direction $\mu$. The external current momentum $q=(\mathbf{q},0)$ has $q_4=0$ by construction from the Feynman-Hellmann approach \cite{Can2020}. $Z_V$ is the RI$^\prime$-MOM renormalisation constant for the local vector current, previously calculated on this lattice in Ref. \cite{Constantinou2014}. A second order Feynman-Hellmann relation for this perturbation has previously been derived in Ref. \cite{Can2025} and reads
\begin{equation}
    \label{eq: 2nd order FH relation}
    \frac{\partial^2 E^{\bm{\lambda}}_{N}(\mathbf{p},\mathbf{q})}{\partial\lambda_1\partial\lambda_2}\bigg|_{\bm{\lambda}=0} = i\frac{T_{\mu\nu}(p,q)-T_{\mu\nu}(p,-q)}{2E_N(\mathbf{p})},
\end{equation}
where $T_{\mu\nu}$ is the Compton amplitude,
\begin{equation}
    T_{\mu\nu}=\rho_{\sigma'\sigma}[T_{\mu\nu}]_{\sigma'\sigma},
\end{equation}
with $[T_{\mu\nu}]_{\sigma'\sigma}$ from Eq. \ref{eq: Compton amplitude}, $\rho=\frac{1}{2}(1+\bm{\sigma}\cdot\bm{s}/m_N)$ being the spin-density matrix, $E^{\bm{\lambda}}_{N}(\mathbf{p},\mathbf{q})$ $(E_N)$ is the (un-) perturbed nucleon energy with momentum $\mathbf{p}$ and is also a function of the momentum carried by the perturbing current, $\mathbf{q}$. In order to isolate the $\mathcal{G}_1$ structure function, we make a certain choice of kinematics, as discussed in Section \ref{sec: Calc details}.

\subsection{Degeneracy of Perturbed Energies}
\label{subsec: diagonalisation}
Equation \eqref{eq: 2nd order FH relation} assumes that there is only a single low-lying perturbed energy, however the polarised Compton amplitude depends on the spin of the nucleon, and hence our perturbing potential will break the degeneracy between the two spin states. Therefore there will be two low-lying perturbed energy states, one corresponding to the spin-up state at $\lambda=0$ and the other to the spin-down state. The spectral form of our perturbed two-point correlator hence has the form
\begin{equation}
    G_\lambda(\mathbf{p},\mathbf{q},t) \approx A_+(\mathbf{p},\mathbf{q})e^{-E^{(+)}_\lambda(\mathbf{p},\mathbf{q})t} + A_-(\mathbf{p},\mathbf{q}) e^{-E^{(-)}_\lambda(\mathbf{p},\mathbf{q}) t},
\end{equation}
with $E^{(\pm)}_\lambda$ being the eigenvalues of the perturbed Hamiltonian, and $A_{\pm}$ being the corresponding overlap factors. As shown in Appendix \ref{app: spin structure}, the perturbed Hamiltonian in the space of the two low-lying states has the following form, up to order $\lambda^2$,
\begin{equation}
    \hat{H}_{\lambda}=\hat{H}_0 + \Delta E_{\text{unpol}}(\lambda_1,\lambda_2)+\lambda_1\lambda_2
    \mathbf{v}\cdot\bm{\sigma}, 
\end{equation}
where $\hat{H}_0|p,s\rangle=E_N(\mathbf{p})|p,s\rangle$ defines the eigenstates of the unperturbed QCD Hamiltonian, the coefficients $\mathbf{v}$ capture the kinematic factors and structure functions from the polarised Compton amplitude, and $\bm{\sigma}=(\sigma_1,\sigma_2,\sigma_3)$ are the Pauli sigma matrices. $\Delta E_{\text{unpol}}(\lambda_1,\lambda_2)$ captures the perturbations from the diagonal, unpolarised components of the Compton amplitude, corresponding to repeated insertions of the $\mathcal{J}_\mu$ or $\mathcal{J}_\nu$ currents. Both have the form
\begin{multline}
\label{eq: mu mu compton amplitude}
    [T_{\mu\mu}(p,q)]_{\sigma'\sigma}=\bigg[-g_{\mu\mu}\mathcal{F}_1(\omega,Q^2)\\
    +\bigg(p_\mu+\frac{p\cdot q}{q^2}q_\mu \bigg)^2\frac{\mathcal{F}_2(\omega,Q^2)}{p\cdot q} \bigg]\delta_{\sigma'\sigma},
\end{multline}
where $\mu$ is just a repeated index and not summed over.  We consider a spatial-temporal component of the polarised Compton amplitude for a nucleon at rest,
\begin{multline}
\label{eq: spatial-temporal polarised CA}
    [T_{i4}(p,q)]_{\sigma'\sigma}=\\ \frac{2m_N i \epsilon_{ijk}}{Q^2}q_k \sigma^{\sigma'\sigma}_{j}\left(\frac{\mathcal{G}_1(\omega,Q^2)}{\omega} + \frac{\mathcal{G}_2(\omega,Q^2)}{\omega}\right)\bigg|_{\omega=0}.
\end{multline}
Here, we note that the term $\mathcal{G}_2(\omega,Q^2)/\omega|_{\omega=0}$ is proportional to the lowest moment of the $g_2$ structure function. The Burkhardt-Cottingham sum rule \cite{Burkhardt1970} predicts that this moment vanishes at all orders in $Q^2$. However, it is worth noting that the Burkhardt-Cottingham sum rule is not a first-principles relation. With this caveat in mind, for our analysis we will assume that this term is negligible and focus on $\mathcal{G}_1(\omega,Q^2)/\omega|_{\omega=0}$.
Comparing Equations \eqref{eq: mu mu compton amplitude} and \eqref{eq: spatial-temporal polarised CA}, we see that the unpolarised Compton amplitude is diagonal in the nucleon spins, whereas the polarised Compton amplitude mixes spin eigenstates.
\newline

For the Feynman-Hellmann relation, Equation \eqref{eq: 2nd order FH relation}, to be valid, we require the perturbed Hamiltonian to be diagonal in spin-space. For certain kinematic choices, as outlined in Appendix \ref{app: spin structure}, the contribution from the polarised Compton amplitude remains diagonal in spin-space. As an example, for the spatial-temporal components of the polarised Compton amplitude, terms proportional to $\sigma_3$ remain diagonal. We therefore select the component $T_{14}(p,q)$ of the Compton amplitude and work with transfer momentum of the form $\mathbf{q}=(0,q_2,0)$ to keep the perturbed Hamiltonian diagonal. This allows us to use a single exponential to extract the energy shift and therefore access the polarised Compton amplitude. As a proof of concept, we also consider one transfer momentum which introduces off-diagonal terms to the Hamiltonian and explore the use of a generalised eigenvalue problem (GEVP) to extract the two low-lying energy states. These two cases are discussed separately in Section \ref{sec: Calc details}.

\section{Calculation and Analysis Details}
\label{sec: Calc details}
The gauge configurations used in this work were generated by the QCDSF Collaboration with 2+1 flavours of dynamical light and strange quarks, using a stout-smeared, non-perturbatively $\mathcal{O}(a)$-improved Wilson action together with a tree-level Symanzik-improved gauge action \cite{Cundy2009}. A single ensemble is analysed, with volume $V=48^3\times96$ and coupling $\beta=5.65$, corresponding to a lattice spacing of $a\simeq 0.068,\mathrm{fm}$ \cite{Bornyakov2015}.
The ensemble lies at the SU(3)-flavour-symmetric point, with the degenerate $u$, $d$, and $s$ quark masses set equal to approximately the physical flavour-singlet mass, $\overline{m}=(2m_s+m_l)/3$ \cite{Bietenholz2010,Bietenholz2011}, yielding $m_\pi\approx412,\mathrm{MeV}$. Further details of the ensemble are given in Table~\ref{tab: lattice details}.
\newline

The primary target of this investigation is the determination of the Bjorken sum rule over a broad range of momentum scales, $0.5 < Q^2 < 8~\mathrm{GeV}^2$, allowing its $Q^2$ dependence to be explored. The specific kinematics of the momentum transfer are displayed in Table~\ref{tab: CA q details}. Since the present analysis focuses on the lowest moment, only the Compton amplitude at $\omega=0$ is required. This allows all calculations to be performed with the nucleon at rest, $\mathbf{p}=(0,0,0)$.
\newline
\begin{table*}[t!]
    %\centering
    \caption{Details of the gauge ensemble used in this work. $N_f$ is the number of fermion flavours, $\beta$ is the inverse gauge coupling, $a$ is the lattice spacing, $c_{SW}$ is the clover coefficient, $\kappa_l$ ($\kappa_s$) is the hopping parameter for the light (strange) quarks, $L$ ($T$) is the spatial (temporal) extent of the lattice,  $m_\pi$ ($m_K$) is the pion (kaon) mass, $Z_V$ is the vector current renormalisation constant on this lattice, and $N_{\rm{cfg}}$ is the number of configurations used.}
    \label{tab: lattice details}
    \begin{tabular}{c c c c c c c c c}
         \hline\hline
         $N_f$ & $\beta$ & $a$ [fm] & $c_{SW}$ & $\kappa_\ell= \kappa_s$ & $L^3\times T$ & $m_\pi=m_K$ [MeV] & $Z_V$ & $N_{\rm{cfg}}$ \\
         \hline
         $2+1$ & $5.65$ & $0.068$ & $2.48$ & $0.122005$ & $48^3\times 96$ & $412$ & $0.8615$ & $535$\\
         \hline
    \end{tabular}
\end{table*}
\begin{table}[t!]
    %\centering
    \caption{Momentum insertions $\mathbf{q}=(q_1,q_2,q_3)$ considered in this work, noting that all $\mathbf{q}$ values have $q_1=0$. Momenta for which we perform a multi-exponential fit are given in italics, and momenta requiring a GEVP have $q_3\neq 0$.}
    \label{tab: CA q details}
    \begin{tabular}{c c}
    \hline\hline
        $\mathbf{q}$ $[2\pi/L]$ & $Q^2$ [GeV$^2$]  \\
        \hline
         \textit{(0,2,0)} & 0.577 \\
         (0,3,0) & 1.298 \\
         (0,4,0) & 2.308 \\
         (0,5,0) & 3.606 \\
         (0,3,5) & 4.904 \\
         (0,6,0) & 5.193 \\
         (0,7,0) & 7.068\\
         \hline\hline
         \end{tabular}
\end{table}

At large Euclidean times, the perturbed two-point correlation function takes the form
\begin{equation}
G_\lambda(\mathbf{p},\mathbf{q},t)\xrightarrow{t\gg0}A_\lambda(\mathbf{p},\mathbf{q})e^{-E^\lambda_{N}(\mathbf{p},\mathbf{q}) t},
\end{equation}
where $E^\lambda_{N}$ is the perturbed energy of the ground state nucleon in the presence of a background field, and $A_\lambda$ is the corresponding overlap factor. For $Q^2 \lesssim 1$ GeV$^2$, excited states are not sufficiently suppressed in the time range over which we extract the perturbed energy shifts. We therefore use a two-exponential ansatz,
\begin{equation}
    G^{(\pm)}(\mathbf{p},t)=A_0^{(\pm)}e^{-E^{(\pm)}_0t} + A_1^{(\pm)}e^{-E^{(\pm)}_1t},
\end{equation}
where $E_0$, $E_1$ refer to the ground and first excited state energies, and $A_0$ and $A_1$ are the corresponding overlaps, with the superscripts $(\pm)$ denoting either the spin-up $(+)$ or spin-down $(-)$ correlator. The required energy splitting is then given by
\begin{equation}
    2\Delta E_{\rm{pol}}=E_0^{(+)}-E_0^{(-)}.
\end{equation}
Further discussion of this approach and a comparison to the ratio method can be found in Appendix A of Ref. \cite{Can2025}. 
\newline

In order to isolate the perturbed energy shifts, we define ratios of components of the positive-parity spin matrix, corresponding to the upper left $2\times2$ sub-block of the two-point correlation function in the Dirac representation. This matrix is denoted as 
\begin{equation}
\label{eq: corr matrix}
    G(\mathbf{p},t)=\begin{pmatrix}
        G^{\uparrow\uparrow}(\mathbf{p},t) & G^{\uparrow\downarrow}(\mathbf{p},t) \\
        G^{\downarrow\uparrow}(\mathbf{p},t) & G^{\downarrow\downarrow}(\mathbf{p},t)
    \end{pmatrix}.
\end{equation}
The treatment of this correlator matrix varies depending on whether the perturbed Hamiltonian is diagonal in the nucleon spins, or non-diagonal. We detail each of these cases below.

\subsection{Diagonal Perturbations}
\label{subsec: diagonal analysis}
\begin{figure}[t!]
    \centering
    \includegraphics[width=\linewidth]{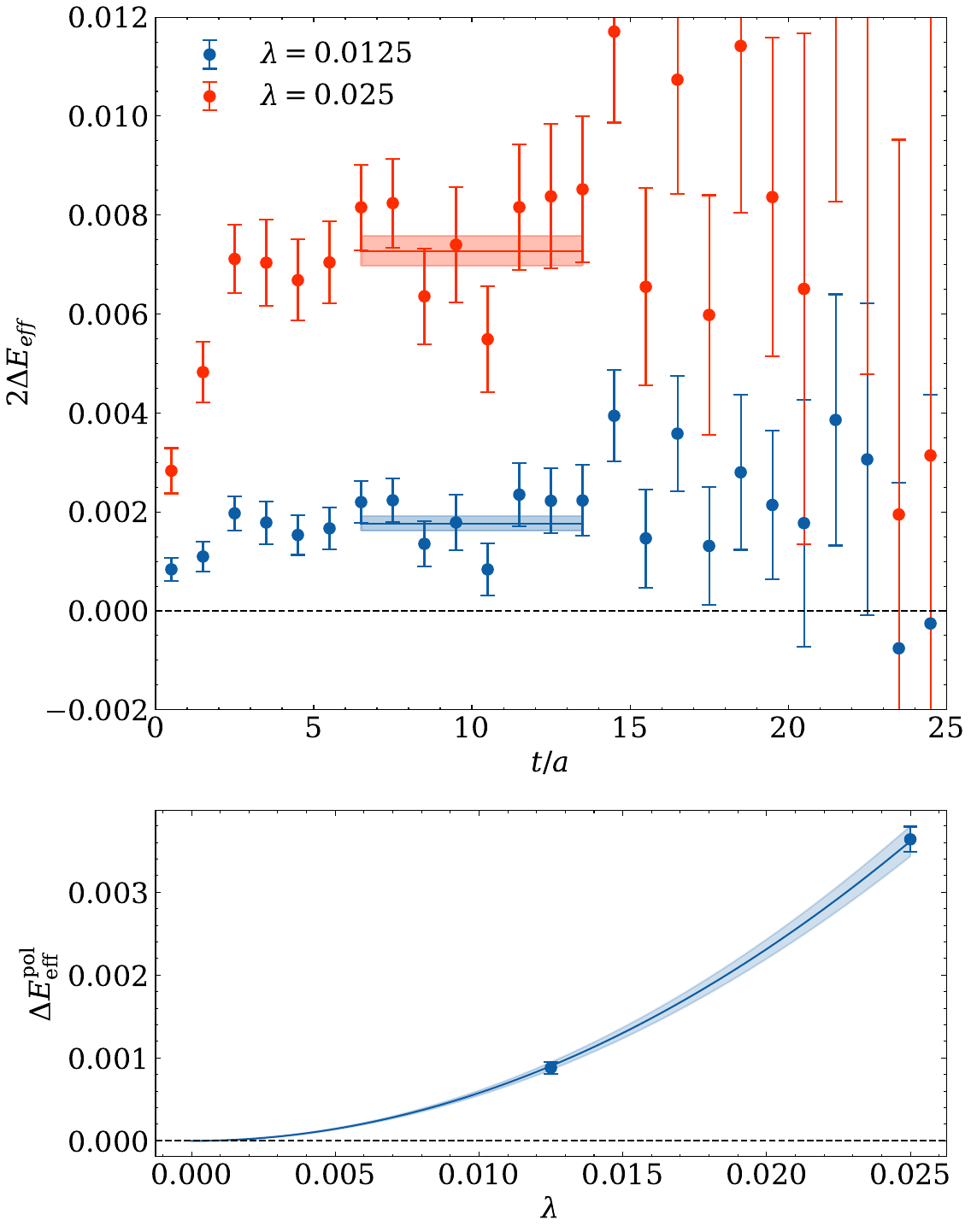}
    \caption{(Top) Effective energy plot analogue for the ratio, Eq. \eqref{eq: diag ratio}, with $\lambda_1 = \lambda_2=\lambda$, for $\lambda=0.0125,0.025$. The calculated energy shifts, uncertainties and their corresponding fit regions are indicated by the horizontal lines and shaded regions, respectively. In this figure, we show results for the $uu$ contribution for $\mathbf{q}=(0,5,0)\frac{2\pi}{L}$ at $Q^2\approx3.6$ GeV$^2$. (Bottom) Corresponding $\lambda$ dependence of the polarised energy shift.}
    \label{fig: diag ratio}
\end{figure}

Diagonal perturbations correspond to the case where the Compton amplitude is proportional to $\sigma_3$. From Equation \eqref{eq: spatial-temporal polarised CA} and considering the component $T_{14}$, this is equivalent to the kinematic choice $q_3=0$. Therefore the spin-up and spin-down states along the $\hat{z}$ axis form an eigenbasis of the perturbed Hamiltonian. In this case, the energy shift due to the polarised Compton amplitude can be isolated through the ratio
\begin{equation}
\label{eq: diag ratio}
    \mathcal{R}(t)=\frac{G^{\uparrow\uparrow}_\lambda(\mathbf{0},t)}{G_0^{\uparrow\uparrow}(\mathbf{0},t)}\frac{G_0^{\downarrow\downarrow}(\mathbf{0},t)}{G^{\downarrow\downarrow}_\lambda(\mathbf{0},t)} \xrightarrow{t\gg0} A_\lambda e^{-2\Delta E_{\rm{pol}}t},
\end{equation}
where the subscripts $\lambda$ and $0$ indicate perturbed and unperturbed correlators respectively and the energy shift due to the polarised Compton amplitude, neglecting terms higher order in $\lambda_1, \lambda_2$ is
\begin{equation}
\label{eq: diag energy shift}
\Delta E_{\rm{pol}} = \lambda_1 \lambda_2 \frac{2q_2}{Q^2}\frac{\mathcal{G}_1(\omega, Q^2)}{\omega}\bigg|_{\omega = 0}.
\end{equation}
The inclusion of the unperturbed correlators in the ratio \eqref{eq: diag ratio} enhances the cancellation of correlated gauge noise between bootstrap samples. A representative plot of this ratio for a nucleon with $\mathbf{q}=(0,5,0)\frac{2\pi}{L}$ is shown in Figure \ref{fig: diag ratio}. The signal region appears around $t=6$, however it is also lost to noise quickly, around $t=14$. Figure \ref{fig: diag ratio} also shows the $\lambda_1=\lambda_2=\lambda$ dependence of the energy shifts and the corresponding quadratic fit. From Equation \eqref{eq: diag energy shift}, the quadratic fit parameter is directly proportional to the Compton structure function $\mathcal{G}_1(\omega, Q^2)$.
\newline

\subsection{Non-Diagonal Perturbations}
\label{subsec: non-diag analysis}
\begin{figure}
    \centering
    \includegraphics[width=\linewidth]{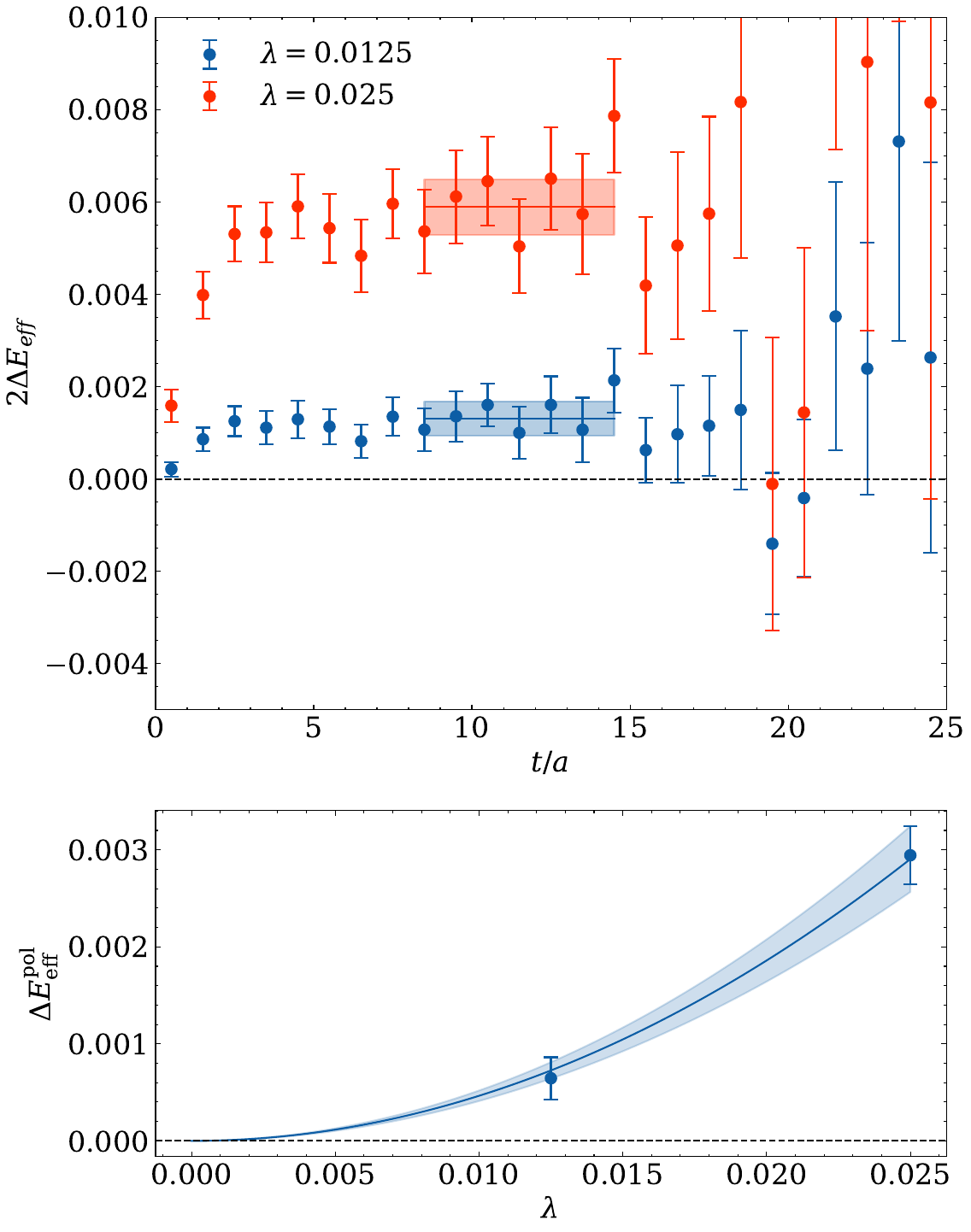}
    \caption{(Top) Effective energy plot analogue for the ratio, Eq. \eqref{eq: non-diag ratio}, with $\lambda_1=\lambda_2=\lambda$, for $\lambda = 0.0125,0.025$. The calculated energy shifts, uncertainties and their corresponding fit regions are indicated by the horizontal lines and shaded regions respectively. In this figure, we show results for the $uu$ contribution for $\mathbf{q}=(0,3,5)\frac{2\pi}{L}$ at $Q^2\sim 4.9$ GeV$^2$. (Bottom) Corresponding $\lambda$ dependence of the polarised energy shift.}
    \label{fig: non-diag eff mass}
\end{figure}

For the case where $q_3\neq 0$, the perturbation is no longer solely proportional to $\sigma_3$ and introduces a mixing between spin components. We consider one such transfer momentum, $\mathbf{q}=(0,3,5)\frac{2\pi}{L}$, as a proof of concept. Again, we use a nucleon momentum of $\mathbf{p}=(0,0,0)$ corresponding to $\omega = 0$. To determine the energy eigenvalues of the perturbed Hamiltonian, we must first diagonalise the correlator matrix \eqref{eq: corr matrix}. We implement a generalised eigenvalue problem (GEVP) to determine the required left $v^{(i)}$ and right $u^{(i)}$ eigenvectors for each value of $\lambda$,
\begin{equation}
    \begin{split}
        G_\lambda(\mathbf{p},t_0+\Delta t)u^{(i)}(t_0,\Delta t)&=c^{(i)}(\Delta t) G_\lambda(\mathbf{p},t_0)u^{(i)}(t_0,\Delta t)\\
        v^{(i)\dagger}(t_0,\Delta t)G_\lambda(\mathbf{p},t_0+\Delta t) &= c^{(i)}(\Delta t)v^{(i)\dagger}(t_0,\Delta t)G_\lambda(\mathbf{p},t_0),
    \end{split}
\end{equation}
for the eigenstates $i=\pm$ and some choice of initial time $t_0$ and separation $\Delta t$, yielding the eigenvalues
\begin{equation}
    c^{(i)}(\Delta t)=e^{-E_\lambda^{(i)}\Delta t},
\end{equation}
with $E_\lambda^{(i)}$ being the perturbed ground state energy of state $i$. The correlator matrix is rotated to the eigenbasis using the left and right generalised eigenvectors,
\begin{equation}
\label{eq: eigenrotation}
    C^{(i)}_\lambda(\mathbf{p},t) = v^{(i)\dagger}C_\lambda(\mathbf{p},t)u^{(i)},
\end{equation}
projecting out the corresponding eigenstate $i=\pm$. To isolate the energy shift from the polarised Compton amplitude, we construct the ratio
\begin{equation}
\label{eq: non-diag ratio}
    \mathcal{R}_\lambda(t) = \frac{C_\lambda^{(+)}(\mathbf{0},t)}{C_0^{(+)}(\mathbf{0},t)}\frac{C_0^{(-)}(\mathbf{0},t)}{C_\lambda^{(-)}(\mathbf{0},t)} \xrightarrow{t\gg0} A_\lambda e^{-2\Delta E_{\rm{pol}}t},
\end{equation}
where $C_0^{(\pm)}(\mathbf{0},t)$ is the rotated unperturbed correlator, using the same eigenvectors as Equation \eqref{eq: eigenrotation}. As with the diagonal case, the inclusion of the unperturbed correlator aids in the cancellation of correlated gauge noise between bootstraps. For the non-diagonal case, the energy shift due to the polarised Compton amplitude, neglecting terms higher order in $\lambda_1,\lambda_2$ is
\begin{equation}
\label{eq: non-diag energy shift}
    \Delta E_{\rm{pol}} = \lambda_1\lambda_2 \frac{2\sqrt{q_2^2 +q_3^2}}{Q^2}\frac{|\mathcal{G}_1(\omega,Q^2)|}{|\omega|}\bigg|_{\omega=0}.
\end{equation}
A representative plot of this ratio is shown in Figure \ref{fig: non-diag eff mass}. As with the diagonal case, the signal region appears early in Euclidean time and is equally lost to noise quite early. It is also worth noting that uncertainty in the calculated ratio is larger than the diagonalised case. This is likely a result of propagating statistical uncertainty in the generalised eigenvectors to the calculation of the energy shifts. Figure \ref{fig: non-diag eff mass} also shows the $\lambda$ dependence of the energy shifts and the corresponding quadratic fit. From Equation \eqref{eq: non-diag energy shift}, the quadratic fit parameter is directly proportional to the Compton structure function $\mathcal{G}_1(\omega,Q^2)$. Correspondingly, it also has a larger uncertainty when compared to the diagonal case.

\subsection{Elastic Contributions}
\label{subsec: elastic contributions}
In the mid $Q^2$ region, it is important to note that both elastic and inelastic processes contribute to the Compton amplitude. Elastic processes are expected to dominate in the low $Q^2$ region, where the higher-twist contributions are also present. Therefore, it is imperative to first quantify the size of the elastic contributions. The elastic pole of the $g_1$ structure function for a quark of flavour $f$ is given by \cite{Hagelstein2015}
\begin{equation}
\label{eq: elastic g1}
    g_{1,f}^{el}(x,Q^2)=\frac{1}{2}F^{(f)}_1(Q^2)G_M^{(f)}(Q^2)\delta(1-x),
\end{equation}
where $G_M(Q^2)$ is the Sachs magnetic form factor,
\begin{equation}
    G_M(Q^2)=F_1(Q^2)+F_2(Q^2),
\end{equation}
with $F_1(Q^2)$ and $F_2(Q^2)$ being the Dirac and Pauli electromagnetic form factors. The $\delta(1-x)$ factor ensures that the only contribution to the dispersion integral, Equation \eqref{eq: dispersion relation}, comes from $x=1$. For a nucleon at rest, we have $\omega = 0$, hence Equation \eqref{eq: elastic g1} gives the elastic contribution to the lowest moment, $\Tilde{M}_1^{(1),\text{el.}}(Q^2)$,
\begin{equation}
    \Tilde{M}_{1,qq}^{(1),el.}(Q^2)=\frac{1}{2}F_1^{(q)}(Q^2)G_M^{(q)}(Q^2).
\end{equation}
The elastic contribution is subtracted from the calculated moments to give the inelastic part of the moment,
\begin{equation}
    \Tilde{M}_{1,qq}^{(1),\text{inel.}}(Q^2)=\Tilde{M}_{1,qq}^{(1)}(Q^2) - \Tilde{M}_{1,qq}^{(1),\text{el.}}(Q^2),
\end{equation}
for $qq=uu,dd$. The electromagnetic form factors $F_1(Q^2)$ and $F_2(Q^2)$ are calculated from three-point function methods on the same set of gauge configurations. We fit both $F_1(Q^2)$ and $G_M(Q^2)$ with a dipole parameterisation, yielding a $1/Q^8$ dependence on the elastic contributions. Further details on the calculation of the electromagnetic form factors and the elastic contributions can be found in Appendix \ref{app: elastic contributions}.

\section{Results and Discussion}
\label{sec: Results}
Before presenting our results, we first comment on the reduction of discretisation errors arising from broken rotational symmetry on the lattice. In Refs.~\cite{Gockeler2006,Schar2025, Schar2025b}, a lattice operator product expansion (LOPE) expression for the Compton amplitude was derived up to first order in the covariant derivative. Starting from the lowest order expression and following the procedure discussed in Section \ref{sec: FH Approach} and Appendix \ref{app: spin structure}, we obtain an equivalent, discretised expression for the Compton amplitude. Matching terms with the continuum version, the LOPE-improvement is achieved by replacing the kinematic factor
\begin{equation}
    \frac{q_i}{Q^2}\to\frac{\sin q_i}{\sum_i [\sin^2 q_i + (1-\cos q_i)^2]},
\end{equation}
in Equations \eqref{eq: diag energy shift} and \eqref{eq: non-diag energy shift}. A similar expression appears in our calculation of the Gross-Llewellyn Smith sum rule \cite{Can2025}, where an alternative derivation of this factor is shown.

\subsection{Flavour Structure of the Moments}
\label{subsec: ellis jaffe}
\begin{figure*}[t!]
    \centering
    \includegraphics[width=\linewidth]{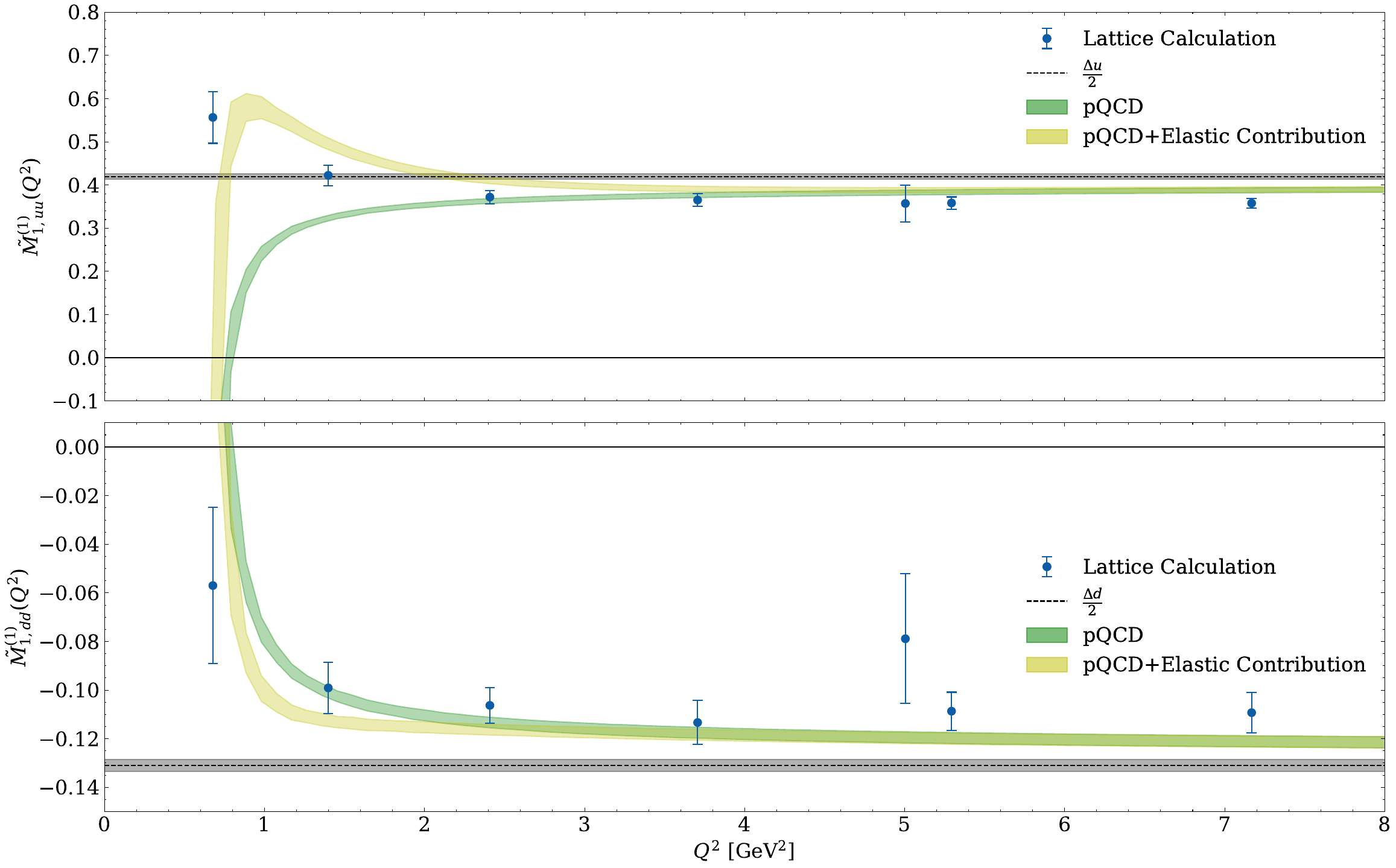}
    \caption{The $uu$ and $dd$ contributions to the lowest moment of the $g_1$ structure function, as a function of $Q^2$. We compare the raw lattice results (blue circles) with the QPM model prediction (grey band), perturbative QCD result at N$^3$LO accuracy (green band) and perturbative QCD result with elastic contributions added (yellow band).}
    \label{fig: quark level moments}
\end{figure*}

We first present our results for the $uu$ and $dd$ contributions to the lowest moment of the $g_1$ structure function, calculated using the procedure outlined in Section \ref{sec: Calc details}
\begin{equation}
    \Tilde{M}^{(1)}_{1,qq}(Q^2) = \frac{\mathcal{G}^{qq}_1(\omega,Q^2)}{4\omega}\bigg|_{\omega=0},
\end{equation}
at each of the $Q^2$ indicated in Table \ref{tab: CA q details}, where $qq$ labels the flavour-diagonal pieces, $uu$ and $dd$. We show the LOPE-improved moments in Figure \ref{fig: quark level moments} for the $uu$ and $dd$ contributions. Our calculated moments cover a wide range of $Q^2$, incorporating both perturbative and non-perturbative scales where different physical phenomena are apparent. We therefore divide the discussion to two regions, although the boundary between these two regions is not definitive and likely exhibit quark mass and lattice spacing dependence. 
\newline

In the large $Q^2$ region, $Q^2 \gtrsim 4$ GeV$^2$, the moments are dominated by the leading-twist parton distributions and their associated contributions from perturbative QCD. In the quark-parton model (QPM), the lowest moment of the $g_1$ structure function is given by
\begin{equation}
\label{eq: qpm ellis jaffe}
    \Tilde{M}_{1,qq}^{(1)}(Q^2)\xrightarrow{Q^2\to\infty} \frac{\Delta q}{2},
\end{equation}
with $\Delta q$ being the fraction of the hadron's spin carried by the quark $q$. At the hadronic level, this defines the Ellis-Jaffe sum rule \cite{Ellis1973}, where the hadronic structure function $g_1$ is the charge-weighted sum of the quark spin contributions. Hence to compare our results to the QPM, we also calculate $\Delta q$ for both $u$ and $d$ quarks on the same set of configurations using three-point function methods. Our values for $\Delta q$ and associated statistical uncertainty are shown as grey bands in Figure \ref{fig: quark level moments}. We utilise three source-sink separations and a two-exponential ansatz for both the two- and three-point correlation functions, and renormalise the axial current in the RI$^\prime$-MOM scheme \cite{Constantinou2014}. As a rigorous evaluation of these matrix elements is not the focus of this work, we leave a detailed discussion of their calculation to Appendix \ref{app: gA}. We see at our largest $Q^2$ point that our calculated moments for both $u$ and $d$ quarks still deviate from the QPM prediction.
\newline

We also compare our calculated moments to the most precise perturbative QCD calculations. In full QCD, the QPM receives perturbative corrections which have been calculated to N$^3$LO \cite{Kataev1994,Kataev2005,Baikov2010},
\begin{equation}
\label{eq: pert ellis jaffe}
    \Tilde{M}_{1,qq}^{(1)}(Q^2)=\frac{\Delta q}{2}\bigg[1 - \sum_{i=1}^4 C_i a^i_s(Q^2) \bigg] + \frac{\Delta_{HT}}{Q^2} + \ldots
\end{equation}
where the Wilson coefficients are $C_1 = 1$, $C_2 = \frac{55}{12}-\frac{1}{3}n_f$, $C_3 = 41.440 - 8.020n_f +0.177n_f^2$ and $C_4=479.4-117.6n_f+7.46n_f^2-0.1037n_f^3$, with $n_f=3$, $a_s(Q^2)\equiv \alpha^{\overline{\rm{MS}}}_s(Q^2)/\pi$ and $\Delta_{HT}$ denotes the sub-leading twist contributions to the sum rule. The solid green band in Figure \ref{fig: quark level moments} shows the leading-twist result with perturbative corrections, given by Eq. \eqref{eq: pert ellis jaffe} without the higher-twist term, calculated using $\alpha_s(Q^2)$ estimated to 4-loop accuracy in the $\overline{\text{MS}}$ scheme following the PDG definition \cite{PDG2024} and with the three flavour $\Lambda_{\overline{\text{MS}}}^{(3)}=338(10)$ MeV quoted by the Flavour Lattice Averaging Group \cite{FLAG2024}, based on determinations from Refs. \cite{Maltman2008, PACS-CS2009, McNeile2010, Chakraborty2014, Bruno2017, Bazavov2019, Ayala2020, Cali2020, Petreczky2020}. The running of $\alpha_s$ is performed using the $\texttt{CRunDec}$ package \cite{Chetyrkin2000, Herren2017}. Our calculated moments are in good agreement with the perturbatively-evolved, leading-twist expectation in this high $Q^2$ region, albeit with mild discretisation artifacts, as was seen in our similar study of the Gross-Llewellyn Smith sum rule \cite{Can2025}. The calculated moment at $Q^2\sim5$ GeV$^2$ corresponds to the single non-diagonal kinematics point and has a correspondingly larger uncertainty in the moment.
\newline

The mid $Q^2$ region, $0.5\lesssim Q^2 \lesssim 4$ GeV$^2$, is significant for investigating higher-twist contributions as the power corrections go as $Q^{-2}$. However simultaneously in this region, the $g_1$ structure function becomes dominated by the elastic pole, as discussed in Section \ref{subsec: elastic contributions}. Indeed, we see the lattice calculations deviate from the perturbative prediction for the up quark as $Q^2$ decreases. Upon addition of the elastic contribution to the perturbative result, shown by the solid yellow bands in Figure \ref{fig: quark level moments}, we begin to capture this departure from the pure perturbative result. The discrepancy between the calculated moments and yellow bands is indicative of higher-twist contributions, however given some deviation is also present at large $Q^2$, full control over discretisation artifacts is necessary before a rigorous calculation of the higher-twist term. With our limited kinematic choices, we are only able to calculate moments with $\mathbf{q}=(0,q_2,0)$, with our lowest transfer momentum being $\mathbf{q}=(0,2,0)\times\frac{2\pi}{L}$. Higher resolution in $Q^2$, particularly in this mid $Q^2$ region, would further assist in the extraction of the higher-twist term.

\subsection{The Bjorken Sum Rule}
\label{subsec: bjorken}
\begin{figure*}[t!]
    \centering
    \includegraphics[width=\linewidth]{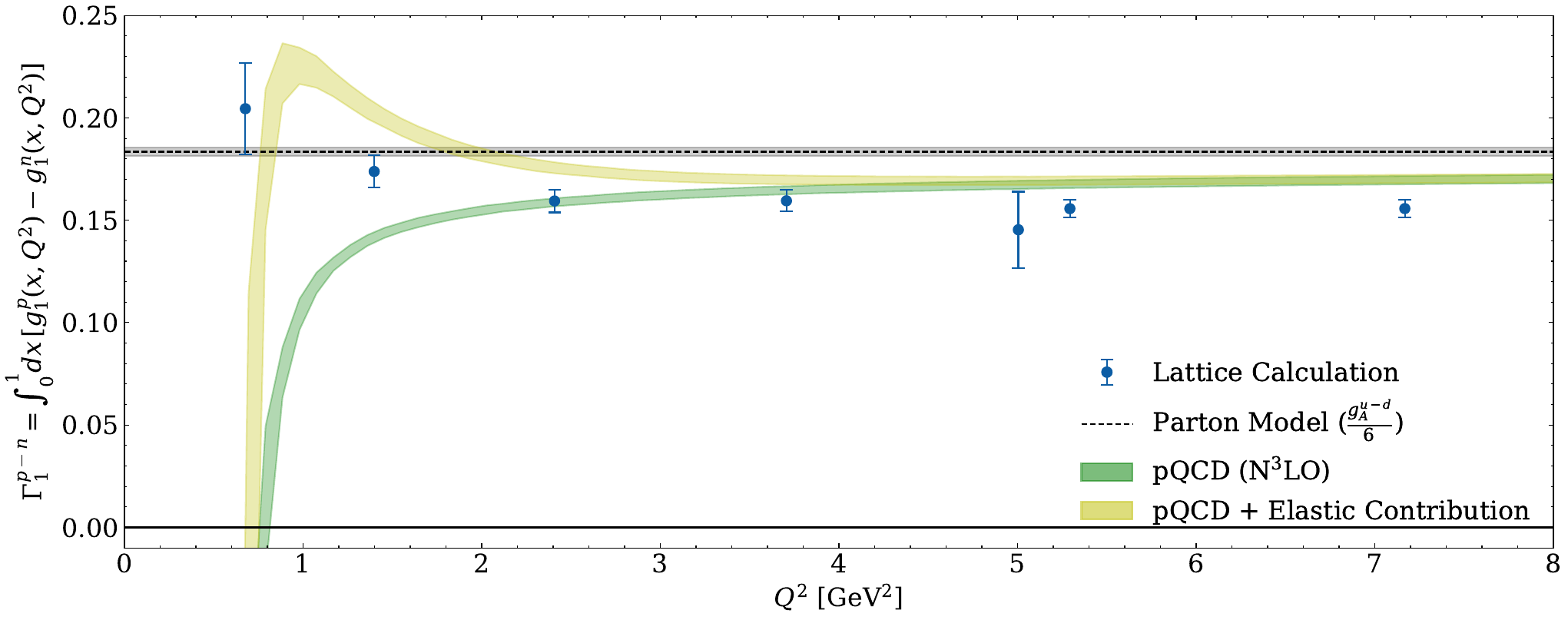}
    \caption{Lattice QCD calculation of the Bjorken sum rule, as a function of $Q^2$. Also plotted is the twist-two prediction for the Bjorken sum rule, $\frac{g_A^{u-d}}{6}$, the N$^3$LO perturbative corrections to the twist-two prediction and N$^3$LO perturbative QCD result plus elastic contribution. $g_A^{u-d}$ is calculated on the same set of gauge configurations.}
    \label{fig: Bjorken sum rule}
\end{figure*}

Recall the Bjorken sum rule is given by
\begin{equation}
    \Gamma_1^{p-n}(Q^2) = \int_0^1dx\,[g_1^p(x,Q^2)-g_1^n(x,Q^2)],
\end{equation}
and is related to the calculated quark-level moments by
\begin{equation}
\label{eq: quark to bjorken}
    \Gamma^{p-n}_1(Q^2) = \frac{1}{3}[\Tilde{M}^{(1)}_{1,uu}(Q^2) - \Tilde{M}_{1,dd}^{(1)}(Q^2)].
\end{equation}
We show our LOPE-improved lattice results in Figure \ref{fig: Bjorken sum rule}, compared with the QPM prediction, perturbative QCD and elastic contributions. As with the quark-level results, we divide our discussion into two regions in $Q^2$.
\newline

In the high $Q^2$ region, $Q^2 \gtrsim 4$ GeV$^2$, the moments are dominated by the leading-twist parton distributions and their associated corrections from perturbative QCD. Combining Equations \eqref{eq: qpm ellis jaffe} and \eqref{eq: quark to bjorken}, we have the quark parton model (QPM) prediction for the Bjorken sum rule,
\begin{equation}
    \Gamma^{p-n,\rm{QPM}}_1(Q^2)\xrightarrow{Q^2\to\infty}\frac{g_A^{u-d}}{6}.
\end{equation}
where we have identified the isovector axial charge $g_A^{u-d}=\Delta u - \Delta d$. We plot our calculated $g_A^{u-d}$ along with its statistical uncertainty as the grey band in Figure \ref{fig: Bjorken sum rule}. We note that our calculation of $g_A^{u-d}$ with three-point function methods is consistent with a previous Feynman-Hellmann calculation on the same lattice \cite{QCDSFUKQCDCSSM2023}. We see our calculated moments lie below the QPM prediction, even at our largest $Q^2$ value. Applying the perturbative corrections to the QPM, we have the perturbative QCD prediction for the Bjorken sum rule, 
\begin{equation}
\label{eq: pert Bjorken sum rule}
    \Gamma_1^{p-n}(Q^2)=\frac{g_A^{u-d}}{6}\bigg[1 - \sum_{i=1}^4 C_i a^i_s(Q^2) \bigg] + \frac{\Delta_{HT}}{Q^2} + \ldots
\end{equation}
where the $C_i$ terms are the same as in Equation \eqref{eq: pert ellis jaffe}. The perturbative prediction is plotted as the solid green band in Figure \ref{fig: Bjorken sum rule}. We find our moments follow the qualitative behaviour of the perturbative result at large $Q^2$, however are offset due to discretisation effects. In principle, the calculated moments can be used to extract an effective strong coupling in this large $Q^2$ region \cite{Deur2005, Deur2016, Deur2023}. As both $g_A^{u-d}$ and $\Gamma^{p-n}$ are calculable with lattice QCD, the deviation between these two quantities can be used to define an effective strong coupling, $\alpha_s^{\text{eff}}$, derived from a hadronic quantity. We leave such a study for future work, after a continuum and chiral extrapolation of the lattice results.

In the mid $Q^2$ region, $0.5$ GeV$^2$ $\lesssim Q^2 \lesssim 4$ GeV$^2$, we begin to see the impact of the elastic contributions, causing the lattice moments to increase rapidly with decreasing $Q^2$. Upon addition of the elastic contribution to the perturbative curve, we see the lattice moments more closely follow this curve, however with some offset. The offset in this region indicates the presence of higher-twist corrections, however given the discrepancy at large $Q^2$, discretisation artifacts likely also contribute. This motivates further investigation into the lattice-spacing dependence of our calculated moments, which we leave as future work. With discretisation artifacts under control, higher-twist corrections, $\Delta_{HT}$, could then be reliably estimated. Model calculations \cite{Balla1997, Balitsky1989, Stein1995, Ehrnsperger1993} and experimental results \cite{Ji1997, Yu2021, Cocuzza2025} suggest these higher-twist corrections are small. It is worth noting that the calculation of higher-twist effects is intimately linked to the divergence of the perturbative series, and that renormalon ambiguity resembling $\Lambda_{QCD}^2/Q^2$ behaviour \cite{Mueller1993, Martinelli1996, Beneke2000, Braun2022} further complicates a rigorous estimate of $\Delta_{HT}$.
\newline

\section{Summary}
\label{sec: Summary}
In this work, we have performed a lattice QCD calculation of the first moment of the nucleon's polarised $g_1$ structure function via a Feynman-Hellmann calculation of the polarised forward Compton amplitude. Our calculations were performed on a single lattice ensemble with volume $48^3\times 96$, lattice spacing $a=0.068$ fm and pion mass tuned to the SU(3) symmetric point, $m_\pi \approx 412$ MeV. Our results cover the $0.5 \lesssim Q^2 \lesssim 8$ GeV$^2$ range, which includes both perturbative and non-perturbative regions. We expand our Feynman-Hellmann analysis to account for perturbations which are not diagonal in the nucleon spins through the use of a GEVP. Additionally, we utilised a LOPE-based parameterisation of the Compton amplitude to reduce the impact of discretisation artifacts on our results. We assessed the agreement of our moments calculated non-perturbatively on the lattice with the perturbative evolution of the leading-twist predictions for the Bjorken sum rule. Our results show deviation from the perturbative QCD predictions in the low $Q^2$ region, hinting at non-negligible higher-twist contributions. A calculation with improved precision, together with several lattice spacings and lighter quark masses, would permit a study of the $1/Q^2$ power corrections after subtraction of the elastic contribution. A direct determination of $\alpha_s$ from the Bjorken sum rule will require a continuum and quark-mass extrapolation together with a controlled treatment of higher-twist and other finite-$Q^2$ effects. The method presented in this work is immediately generalisable to finite nucleon momentum, allowing for the full $\omega$ dependence of the polarised structure functions to be calculated. From this $\omega$ dependence, higher moments of the structure functions could be accessed. Nevertheless, this work demonstrates the feasibility of determining a non-singlet DIS moment from the lattice using the Feynman--Hellmann method.

\begin{acknowledgments}
The numerical configuration generation (using the BQCD lattice QCD program \cite{Haar2017}) and data analysis (using the Chroma software library \cite{Edwards2004}) was carried out on the Extreme Scaling Service (DiRAC, EPCC, Edinburgh, UK), the Data Intensive Service (DiRAC, CSD3, Cambridge, UK), the Gauss Centre for Supercomputing (NIC, J\"ulich, Germany), the NHR Alliance (Germany), and resources provided by the NCI National Facility in Canberra, Australia (supported by the Australian Commonwealth Government), the Pawsey Supercomputing Centre (supported by the Australian Commonwealth Government and the Government of Western Australia), and the Phoenix HPC service (Adelaide University). PELR is supported in part by the STFC under contract ST/G00062X/1. GS is supported by DFG Grant SCHI 179/8-1. JAC and TGS are supported by an Australian Government Research Training Program (RTP) Scholarship. JAC partially received funding from the European Research Council (ERC) via the project ``LEEX” grant agreement 101170304, funded by the European Union. Views and opinions expressed are however those of the author(s) only and do not necessarily reflect those of the European Union or the European Research Council Executive Agency (ERCEA). Neither the European Union nor the ERCEA can be held responsible for them. RDY and JMZ are supported by the Australian Research Council grants DP190100297, DP220103098, and DP240102839.
\end{acknowledgments}

\appendix

\section{Spin Structure of the Compton Amplitude}
\label{app: spin structure}
In this appendix, we provide more details on the diagonalisation of the Compton amplitude and the required kinematic choices for the diagonal and non-diagonal cases. Starting from Hamiltonian perturbation theory, our perturbed action introduces $\lambda$-dependent perturbations to the Hamiltonian,
\begin{equation}
    \hat{H}_\lambda = \hat{H}_0+\lambda\hat{H}_1+\lambda^2\hat{H}_2 + \mathcal{O}(\lambda^3),
\end{equation}
where $\hat{H}_0$ is the unperturbed QCD Hamiltonian, $\hat{H}_0|p,s\rangle=E_N(\mathbf{p})|p,s\rangle$, and the terms higher order in $\lambda$ are related to matrix elements of the perturbing current via the Feynman-Hellmann theorem. As we consider forward matrix elements in this work, terms proportional to $\lambda$ will not contribute, as each insertion of the current carries momentum $\mathbf{q}$. Therefore, our energy splitting appears only at second order in $\lambda$. Using the perturbation to the QCD action described in Equation \eqref{eq: action perturbation}, the perturbed energy levels are given by
\begin{multline}
\label{eq: energy Taylor expansion}
    E^\lambda_N(\mathbf{p},\mathbf{q}) = E_N(\mathbf{p}) + \frac{\lambda_1^2}{2}\frac{\partial^2E_N^\lambda(\mathbf{p},\mathbf{q})}{\partial\lambda_1^2} \\+ \frac{\lambda^2_2}{2}\frac{\partial^2 E_N^\lambda(\mathbf{p},\mathbf{q})}{\partial \lambda_2^2} + \lambda_1\lambda_2 \frac{\partial^2 E_N^\lambda(\mathbf{p},\mathbf{q})}{\partial \lambda_1 \partial \lambda_2}
\end{multline}
Using the Feynman-Hellmann relation for repeated insertions of the same current \cite{Can2020},
\begin{equation}
    \frac{\partial^2 E_N^\lambda(\mathbf{p},\mathbf{q})}{\partial \lambda^2} = -\frac{T_{\mu\mu}(\mathbf{p},\mathbf{q})-T_{\mu\mu}(\mathbf{p},-\mathbf{q})}{2E_N(\mathbf{p})},
\end{equation}
the first two perturbations are proportional to the unpolarised Compton amplitude and are hence diagonal in the space of nucleon spins. In contrast, using Equation \eqref{eq: 2nd order FH relation}, the mixed derivative term is proportional to the polarised Compton amplitude. To isolate $\mathcal{G}_1(\omega,Q^2)$, we choose the indices $\mu=1$ and $\nu=4$ of the Compton amplitude $[T_{\mu\nu}(p,q)]_{\sigma'\sigma}$. From Equation \eqref{eq: Tensor decomposition}, with the additional constraint of considering a nucleon at rest $\mathbf{p}=(0,0,0)$ and noting $q_4=0$ in the Feynman-Hellmann approach, this component of the Compton amplitude in Euclidean space is
\begin{multline}
    [T_{14}(p,q)]_{\sigma'\sigma}= \\
    -\frac{2i}{Q^2}\epsilon_{1ij}q_{i}\sigma_{j}^{\sigma'\sigma}\left( \frac{\mathcal{G}_1(\omega,Q^2)}{\omega}\bigg|_{\omega=0} + \frac{\mathcal{G}_2(\omega,Q^2)}{\omega}\bigg|_{\omega=0}\right),
\end{multline}
where $\epsilon_{123}=+1$. As discussed in the main text, we assume the contribution from the $\mathcal{G}_2(\omega,Q^2)/\omega|_{\omega=0}$ term is zero due to the Burkhardt-Cottingham sum rule. Evidently, by the diagonal nature of $\sigma^3$, this component of the Compton amplitude remains diagonal in the space of nucleon spins for current momentum of the form $\mathbf{q}=(0,q_2,0)$. For more general current momentum, the off-diagonal nature of the $\sigma^1$ and $\sigma^2$ matrices causes a non-trivial mixing of states. The final term of Equation \eqref{eq: energy Taylor expansion} is, with added spin indices,
\begin{equation}
    \bigg[ \frac{\partial^2 E_N^\lambda(\mathbf{p},\mathbf{q})}{\partial \lambda_1 \partial \lambda_2}\bigg]_{\sigma'\sigma} = \frac{2}{Q^2}\epsilon_{1ij}q_{i}\sigma_{j}^{\sigma'\sigma} \frac{\mathcal{G}_1(\omega,Q^2)}{\omega}\bigg|_{\omega=0}.
\end{equation}
We package the kinematic factors and structure function into a coefficient $v_j$,
\begin{equation}
    v_j = \frac{2}{Q^2}\epsilon_{1ij}q_i \frac{\mathcal{G}_1(\omega,Q^2)}{\omega}\bigg|_{\omega=0},
\end{equation}
such that this energy shift can be compactly written as
\begin{equation}
    \bigg[ \frac{\partial^2 E_N^\lambda(\mathbf{p},\mathbf{q})}{\partial \lambda_1 \partial \lambda_2}\bigg]_{\sigma'\sigma} = \mathbf{v}\cdot \bm{\sigma}_{\sigma'\sigma.}
\end{equation}
The perturbed Hamiltonian acting on the space of nucleon spins is
\begin{equation}
    \hat{H}_\lambda = \hat{H}_0 + \Delta E_{\text{unpol}}(\lambda_1,\lambda_2)\delta_{\sigma'\sigma} + \lambda_1 \lambda_2 \mathbf{v}\cdot\bm{\sigma}_{\sigma'\sigma},
\end{equation}
where $\Delta E_{\text{unpol}}(\lambda_1,\lambda_2)\delta_{\sigma'\sigma}$ captures the energy shifts from the unpolarised Compton amplitude. Diagonalising this Hamiltonian to determine the energy eigenvalues, we find
\begin{equation}
    E_\lambda^{(\pm)} = E_N(\mathbf{p}) + \Delta E_\text{unpol}(\lambda_1,\lambda_2) \pm \lambda_1 \lambda_2 \sqrt{\mathbf{v}\cdot\mathbf{v}}.
\end{equation}

\section{Elastic Contributions}
\label{app: elastic contributions}
The elastic contribution to the nucleon polarised structure function $g_1$ is given by
\begin{equation}
    \label{eq: app g1 elastic}
    g_{1,f}^{el}(x,Q^2)=\frac{1}{2}F_1^{(f)}(Q^2)G_M^{(f)}(Q^2)\delta(1-x),
\end{equation}
where $G_M(Q^2)$ is the Sachs magnetic form factor,
\begin{equation}
    G_M(Q^2)=F_1(Q^2)+F_2(Q^2),
\end{equation}
with $F_1(Q^2)$ and $F_2(Q^2)$ being the electromagnetic form factors. Our calculation of the form factors follows the conventional three-point function approach, and we detail the main steps for completeness.
\newline

\begin{figure}
    \centering
    \includegraphics[width=\linewidth]{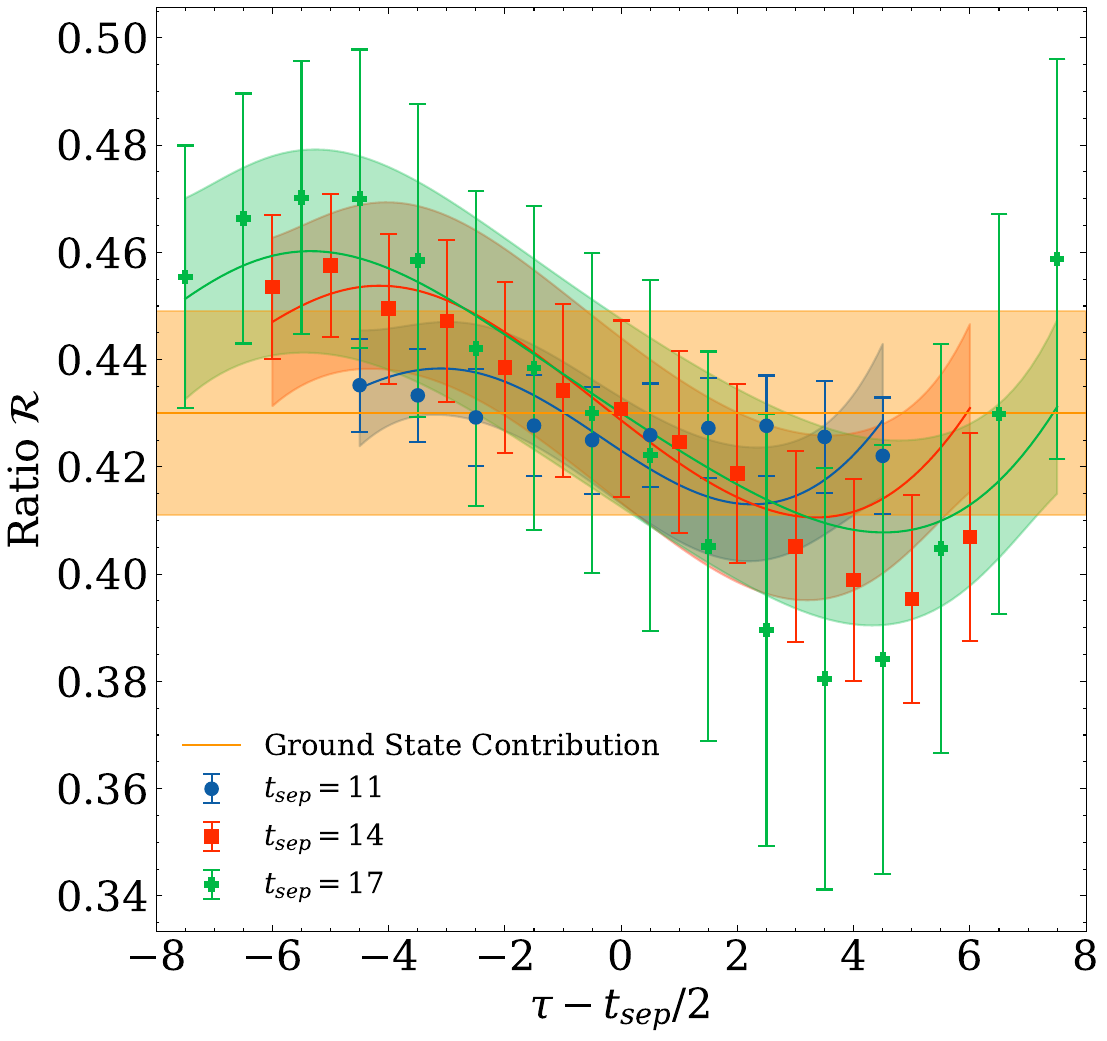}
    \caption{Euclidean time dependence of the ratio in Equation \eqref{eq: ratio def} for three source-sink separations. The inserted operator is $\overline{\psi}\gamma_3\psi$, with a sink momentum $\mathbf{p}'=(1,0,0)$ and source momentum $\mathbf{p}=(2,0,0)$, yielding $Q^2=0.12$ GeV$^2$. The three-point function is projected using the projector $\Gamma_3$, and the resulting ratio is proportional to $G_M$.}
    \label{fig: EM ratio calculation}
\end{figure}

The nucleon matrix element of the electromagnetic current is decomposed in terms of the Dirac ($F_1$) and Pauli ($F_2)$ form factors,
\begin{multline}
    \langle p',s'|\mathcal{J}_\mu(0)|p,s\rangle=\\ \overline{u}(p',s')\bigg[\gamma_\mu F_1(Q^2) + \frac{\sigma_{\mu\nu}q_\nu}{2m_N} F_2(Q^2)\bigg]u(p,s),
\end{multline}
where $u(p,s)$ is a Dirac spinor with 4-momentum $p$ and spin $s$, $q=p'-p$ is the transfer momentum, $Q^2=-q^2$, and $\sigma_{\mu\nu}=\frac{i}{2}[\gamma_\mu,\gamma_\nu]$. The electromagnetic current is given by
\begin{equation}
    \mathcal{J}_\mu = \frac{2}{3}\overline{u}\gamma_\mu u-\frac{1}{3}\overline{d}\gamma_\mu d+\ldots
\end{equation}
The elastic form factors are calculated from ratios of three- and two-point correlation functions. For the Sachs magnetic form factor, we insert the local operators,
\begin{equation}
    j_\mu=\overline{\psi}\gamma_\mu \psi, \hspace{0.5cm} \text{for } \mu=1,2,3,4.
\end{equation}
Matrix elements of these operators are calculated from three-point functions,
\begin{multline}
    G^{(3)}(\mathbf{p}',t';\mathbf{q},\tau;\mathcal{O})=\sum_{\mathbf{x}_1,\mathbf{x}_2}e^{-i\mathbf{p}'\cdot\mathbf{x}_2}e^{i\mathbf{q}\cdot\mathbf{x}_1}\Gamma_{\alpha\beta}\times\\
    \langle \overline{\chi}(\mathbf{x}_2,t')\mathcal{O}(\mathbf{x}_1,\tau)\chi(\mathbf{0},0)\rangle,
\end{multline}
where $\chi$ are the standard nucleon interpolating operators. The local operator $\mathcal{O}$ is inserted at timeslice $\tau$ such that $0<\tau<t'$, where $t'$ is the sink time. $\Gamma$ is a spin-parity projector, with $\Gamma_{\rm{unpol}}=\frac{1}{2}(I+\gamma_4)$ and $\Gamma_i=-\frac{i}{2}(1+\gamma_4)\gamma_i\gamma_5$, with $i=1,2,3$. For this analysis, we neglect disconnected contributions and only consider connected contributions to the three-point function. We construct the following ratio of three- and two-point functions,
\begin{multline}
\label{eq: ratio def}
    \mathcal{R}(\mathbf{p}',t';\mathbf{q},\tau;\mathcal{O})=\frac{G^{(3)}(\mathbf{p}',t';\mathbf{q},\tau;\mathcal{O})}{G^{(2)}(\mathbf{p}',t')}\times\\
    \bigg[\frac{G^{(2)}(\mathbf{p}',t')G^{(2)}(\mathbf{p}',\tau)G^{(2)}(\mathbf{p},t'-\tau)}{G^{(2)}(\mathbf{p},t')G^{(2)}(\mathbf{p},\tau)G^{(2)}(\mathbf{p}',t'-\tau)}\bigg]^{\frac{1}{2}},
\end{multline}
which, assuming ground-state dominance at large Euclidean times, is proportional to the matrix element of the inserted operator. To control excited-state contamination, we use a two-exponential ansatz to model the three-point function when fitting directly to the ratio. The two-point functions are fit beforehand, also using a two-exponential ansatz, to determine the ground-state energy $E_N$ and energy gap $\Delta E_N$,
\begin{equation}
    G^{(2)}(\mathbf{p},t)=A_{2,0}(\mathbf{p})e^{-E_N(\mathbf{p})t}(1 + A_{2,1}(\mathbf{p})e^{-\Delta E_N(\mathbf{p})t}),
\end{equation}
where $A_{2,0}$ and $A_{2,1}$ are overlap factors. The three-point function ansatz is
\begin{multline}
    G^{(3)}(\mathbf{p}',t';\mathbf{q},\tau;\mathcal{O})=A_{3,0}e^{-E_N(\mathbf{p}')(t'-\tau)}e^{-E_N(\mathbf{p})\tau}\times \\
    \big[B_{00} + B_{10}e^{-\Delta E_N(\mathbf{p}')(t'-\tau)} +B_{01}e^{-\Delta E_N(\mathbf{p})\tau}\\ + B_{11}e^{-\Delta E_N(\mathbf{p'})(t'-\tau)}e^{-\Delta E_N(\mathbf{p})\tau}\big],
\end{multline}
where $A_{3,0}$ is an overlap factor which cancels in the ratio \ref{eq: ratio def}, and $B_{00}$ is proportional to the desired matrix element,
\begin{equation}
    B_{00} = \bigg[\frac{E_N(\mathbf{p})E_N(\mathbf{p}')}{(E_N(\mathbf{p})+m_N)(E_N(\mathbf{p}')+m_N)}\bigg]^{\frac{1}{2}}\langle p',s' |j_\mu|p,s\rangle.
\end{equation}
The terms $B_{01}$ and $B_{10}$ are proportional to transition matrix elements between the ground state and first excited state (and vice versa), while $B_{11}$ is proportional to the matrix element between two nucleons in the first excited state. A representative plot of the ratio used to determine $G_M$ is shown in Figure \ref{fig: EM ratio calculation}.
\newline

\begin{figure*}[t]
    \centering
    \includegraphics[width=\linewidth]{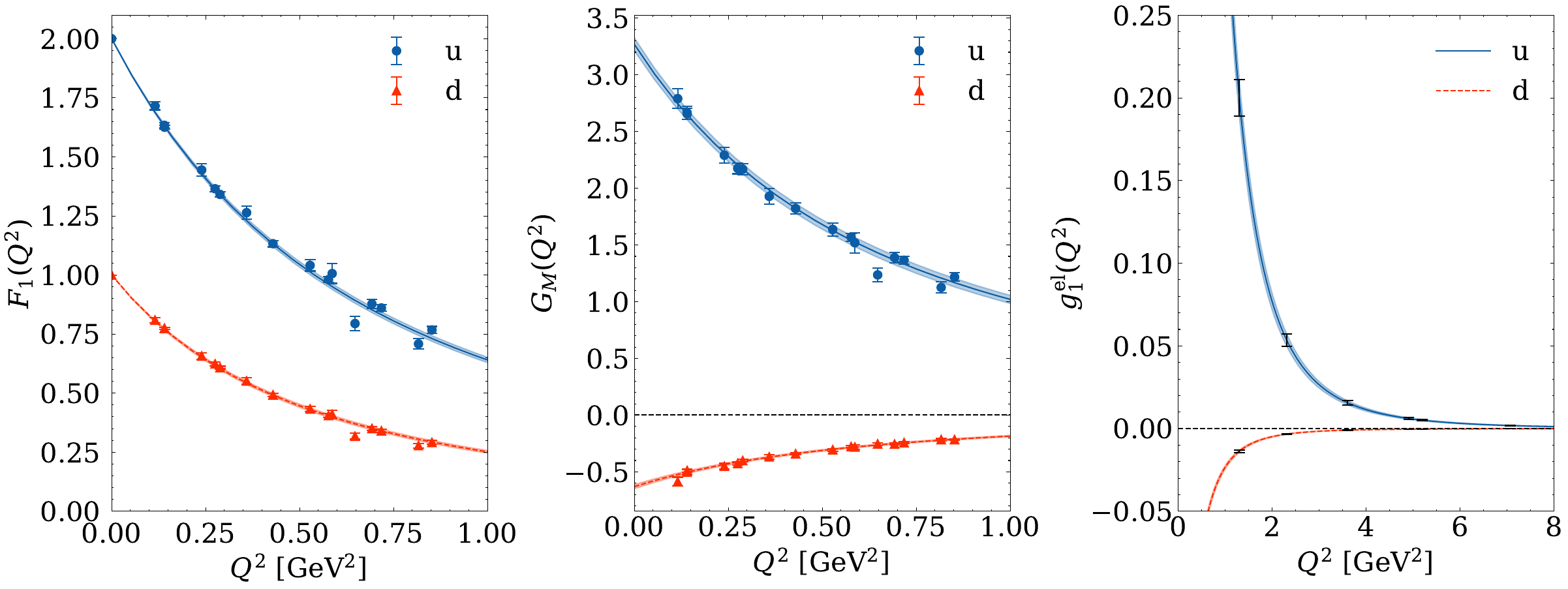}
    \caption{The $Q^2$ dependence of the form factors $F_1(Q^2)$ and $G_M(Q^2)$, as well as the elastic contribution to the $g_1$ structure function (Eq. \eqref{eq: app g1 elastic}). We show results for the $u-$ (blue circles) and $d$-quark (red triangles) form factors, and the $u-$ (blue solid line) and $d$-quark (red dashed line) elastic contribution $g_1^{el}$. $g_1^{el}$ is extrapolated over the whole $Q^2$ range using the dipole parameterisation from Eq. \eqref{eq: dipole parameterisation}.}
    \label{fig: elastic}
\end{figure*}

Repeating this analysis across a range of $Q^2$ values allows us to determine the $Q^2$ dependence of $F_1$ and $G_M$. We renormalise our form factors in the RI$^\prime$-MOM scheme by computing the renormalisation factor
\begin{equation}
    Z^{(u)}_V=\frac{2}{F^{(u)}_1(0)}, \hspace{0.5cm} \text{or}\hspace{0.5cm} Z^{(d)}_V=\frac{1}{F_1^{(d)}(0)},
\end{equation}
for the up and down quarks respectively. We plot the resulting points in Figure \ref{fig: elastic}. We model the $Q^2$ dependence of the form factors using a dipole parameterisation, given by
\begin{equation}
\label{eq: dipole parameterisation}
    F(Q^2)=\frac{F(0)}{\left(1-\frac{Q^2}{\Lambda^2} \right)^2},
\end{equation}
where $F(0)$ and $\Lambda$ are fit parameters. We make use of this parameterisation to extrapolate these form factors beyond the $Q^2$ range accessible via our lattice calculation. In Figure \ref{fig: elastic}, we plot the elastic $g_1$ structure function across the whole $Q^2$ range used in our analysis. For $Q^2\gtrsim4$ GeV$^2$, the elastic contribution to the structure function is practically negligible. 
\newline

\section{Calculation of $g_A^{u-d}$}
\label{app: gA}
The contribution of quark flavour $q$ to the total hadron spin, $\Delta q$, is related to matrix elements of the axial current as follows,
\begin{equation}
    \langle p,s|\overline{\psi}_q(0)\gamma_\mu \gamma_5\psi_q(0)|p,s\rangle = 2s_\mu \Delta q,
\end{equation}
where $s_\mu$ is the spin four-vector. The isovector matrix element $g_A^{u-d}$ is then $g_A^{u-d}=\Delta u-\Delta d$.
\newline

\begin{figure*}[t]
    \centering
        \centering
        \includegraphics[width=0.49\textwidth]{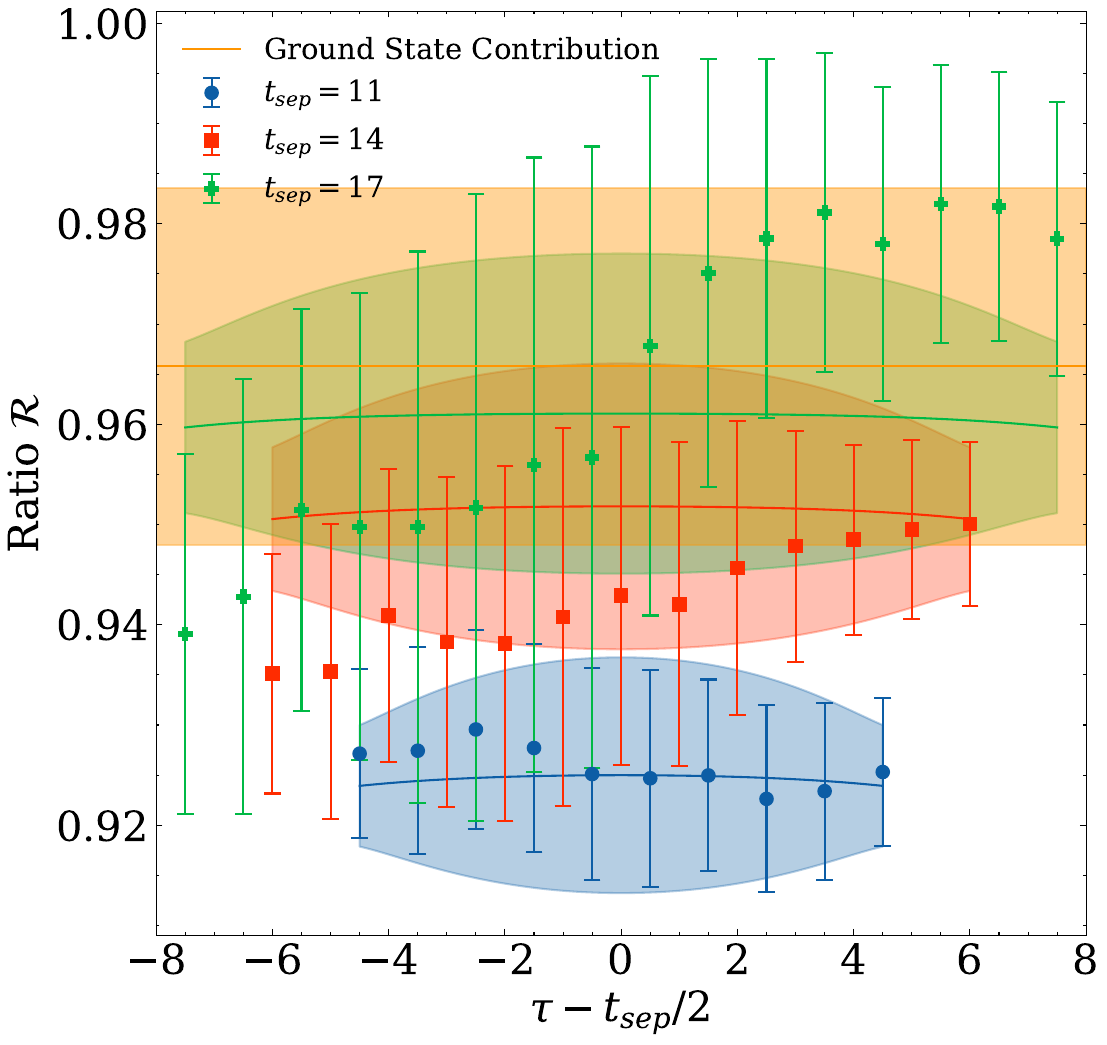}
        \hfill
        \includegraphics[width=0.49\textwidth]{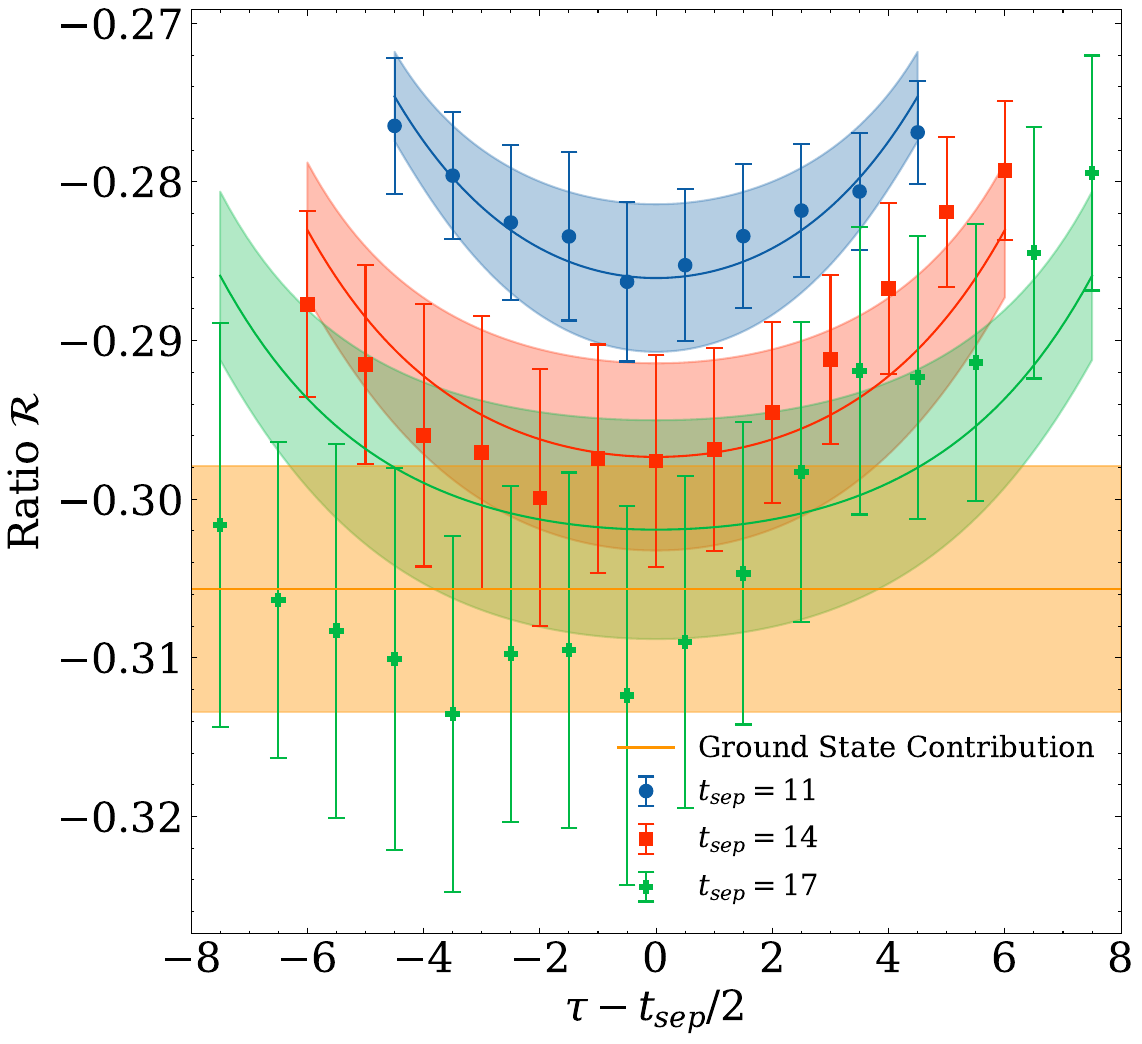}
    \caption{Calculation of $\Delta q$ for both the up (left) and down (right) quarks, using the operator $\overline{\psi}_q(0)\gamma_1\gamma_5\psi_q(0)$ and projector $\Gamma_{\rm{proj}}=\Gamma_{1}$. The horizontal axis is shown in units of $a$.}
    \label{fig: axial ratios}
\end{figure*}

As with the electromagnetic form factors, we compute $g_A^{u-d}$ using three-point function methods. We insert the local operators
\begin{equation}
j_A=\overline{\psi}\gamma_\mu\gamma_5\psi, \hspace{0.5cm} \mu=1,2,3,4,
\end{equation}
and compute the three-point function via sequential inversion through the sink. We consider connected contributions to the three-point function only, and extract the ground state contribution as described in Appendix \ref{app: elastic contributions}. Representative plots of the ratios used to extract $\Delta q_f$ for the up and down quarks are shown in Figure \ref{fig: axial ratios}. This yields the bare results for each quark flavour of $\Delta u=0.958(15)_{\rm{stat}}$ and $\Delta d=-0.299(57)_{\rm{stat}}$. The bare isovector combination is obtained by taking the difference of the individual quark results, yielding $g_A^{u-d}=1.258(15)_{\rm{stat}}$. The RI$^\prime$-MOM renormalisation constant for the axial current has previously been calculated for this lattice and takes the value $Z_A = 0.8754(19)_{\rm{stat}}$ \cite{Constantinou2014}. The renormalised value for $g_A^{u-d}$ is then $g_A^{u-d}=1.101(13)_{\rm{stat}}$, which is the value appearing in the main text.

\bibliography{bjorken}% Produces the bibliography via BibTeX.

\end{document}